\documentclass[prb,aps,superscriptaddress,twocolumn,floats,nofootinbib,showpacs]{revtex4}
\usepackage{amsmath,amssymb,graphicx}
\DeclareMathOperator{\sign}{sgn}

\begin{document}

\title{Dynamics of a one-dimensional spinor Bose liquid: a
  phenomenological approach}

\author{A. Kamenev}
\affiliation{School of Physics and Astronomy, University of
Minnesota, Minneapolis, MN 55455}
\author{L. I. Glazman}
 \affiliation{Department of Physics, Yale
University, P.O. Box 208120,
  New Haven, CT 06520-8120}

\begin{abstract}
  The ground state of a spinor Bose liquid is ferromagnetic, while the softest excitation above the ground state is the magnon mode. The
  dispersion relation of the magnon  in a one-dimensional
  liquid is periodic in the wavenumber $q$ with the period $2\pi n$, determined by the
  density $n$ of the liquid. Dynamic  correlation functions, such as e.g. spin-spin
  correlation function,
  exhibit power-law singularities at the magnon spectrum,
  $\omega\to\omega_m(q,n)$.  Without using any specific model of the
  inter-particle interactions, we relate the corresponding exponents to
  independently  measurable  quantities $\partial\omega_m/\partial q$ and
  $\partial\omega_m/\partial n$.
\end{abstract}

 \maketitle

Bosons with an internal degree of freedom, ``spin'', exhibit a
ferromagnetic ground state \cite{Lieb}. The presence of the internal
states yields an excitation, magnon, in addition to conventional waves
of the mass density \cite{bashkin,lhuillier}. Because of the
ferromagnetic ground-state order, the dispersion relation
$\omega_m(q)$ of magnons at small wave vectors $q$ is
quadratic. In case of spin-isotropic liquid, this is
the softest excitation of the system, since $\omega_m(0)=0$, while the
density waves in the $q\to 0$ limit propagate with a finite sound
velocity $v$.  Away from the isotropic limit $\omega_m(0)>0$, still
magnon is the lowest energy {\em spin excitation}.  One may think of a
magnon as of a ``quantum impurity'' with the spin opposite to the
majority direction, moving in a spin-polarized host liquid.

A spin flip in a compressible system may excite a wave of density in
it, which affects the spin-spin correlation function $A_m(q,\omega)$.
This back-action of the medium is the strongest in a one-dimensional
(1D) system. In a 1D Heisenberg ferromagnet on a rigid lattice,
$A_m(q,\omega)\propto f(q)\delta (\omega-\omega_m(q))$ with $f(q)$
being periodic in the reciprocal lattice. A finite compressibility of
liquid transforms the $\delta$-function response into a power-law edge
singularity, $A_m(q,\omega)\propto (\omega-\omega_m(q))^{\mu_m}\theta
(\omega-\omega_m(q))$, at the magnon spectrum. While the existence of
that singularity can be argued on the basis of the scale invariance
\cite{zvon}, the exponent $\mu_m(q)$ has been evaluated thus far only
for an SU(2) symmetric system in the limit of strong repulsion between
the bosons~\cite{zvon,akirako}, and for an inegrable Yang-Gaudin
model~\cite{zvon1}.

The singularities in $A_m(q,\omega)$, along with the previously studied
singularities in the dynamic responses of single-component quantum
fluids~\cite{khodas,adilet}, and the spectral function singularity in
the quantum impurity problem share common root and can be
related~\cite{pustilnik,adilet1,lamacraft} to the physics of Fermi edge
singularity. The exponent $\mu_m(q)> -1$ originates in the
interaction of the ``quantum impurity'' with low-energy density waves of the
majority spin polarization.

Here we express the exponent $\mu_m(q)$ through the dispersion of the
magnon mode $\omega_m=\omega_m(q,n)$ and its derivatives over the
wavenumber $q$ and the density of the majority spin component $n$.
For this purpose, we find the constants of the ``quantum impurity''
Hamiltonian which describes the dynamics of the magnon excitation and
its interaction with fluctuations of density.  That, in turn, reduces
the problem of finding the critical behavior of a dynamic response to
a much simpler problem of evaluation of an excitation energy as a
function of $n$ and $q$. Our results for $\mu_m(q)$ are not limited to
small momenta and independent of the interaction strength and the
presence of the SU(2) symmetry. In addition to $A_m(q,\omega)$, we
consider also the single-particle spectral function $A_d(q,\omega)$ of
a boson with the spin opposite to the majority polarization added to
the fully polarized liquid. It also displays a power-law threshold
behavior, and we find the corresponding wavenumber-dependent exponent
$\mu_d(q)$.

To find the exponents $\mu_m(q)$ and $\mu_d(q)$ we start with introducing a
representation suitable for the description of the low-energy spin
dynamics. For definiteness we consider spin-$1/2$ particles, and
assume all spins point in the positive-$z$ direction in the state with
maximal spin polarization. Addressing the spin density operator,
\begin{equation}
{\bf s}(x)=\frac{1}{2}\sum_j {\boldmath\sigma}^{(j)}\delta (x-x_j),
\label{spindensity}
\end{equation}
we concentrate first on its $z$-component. Within the set of states of
maximal spin polarization, spin density fluctuations are proportional
to the fluctuations of the number density. Thus we represent the
long-wavelength part of density $s_z(x)$ in terms of boson field
$\phi$ as $s_z(x)=\partial_x\phi/(2\pi)$, where field $\phi$ has the
meaning of the displacement field of spin-up particles.  The
long-wavelength dynamics of the displacements is described by the
Luttinger liquid Hamiltonian\cite{Haldane},
\begin{equation}
H_0=\frac{v}{2\pi}
\int dx\left[
K\left(\partial_x\theta \right)^2
+\frac{1}{K}\left(\partial_x\phi\right)^2\right],
\label{liquid}
\end{equation}
where $\theta$ is the conjugate to $\phi$ variable,
$[\phi(x),\partial_x\theta(y)]=i\pi\delta(x-y)$. For Galilean
invariant system the Luttinger parameter is $K=\pi n/(m_\uparrow v)$
with $m_\uparrow$ being the bare mass of spin-up bosons.

Considering the operator $s_-(x)$ in the sub-space of excitations with
energies close to the magnon energy $\omega_m(q)$, we replace the general form
Eq.~(\ref{spindensity}) by an operator
\begin{equation}
s_-(x)\propto d^\dagger(x)e^{iqx}e^{i\theta(x)}.
\label{sminus}
\end{equation}
Acting on a fully polarized state, this operator ``extracts'' a
particle with spin-up, and replaces it (by acting with the creation
operator $d^\dagger$ on vacuum) with a spin-down particle at the same
point $x$.
The representation
Eq.~(\ref{sminus}) of $s_-(x)$ operator
is adequate
for $|q|\leq\pi n$, and requires~\cite{Haldane} a generalization
(considered later in this paper) for larger momenta.

The Hamiltonian describing the magnon dynamics should
include the dynamics of the $d$-quasiparticle and its interaction with the
fields $\phi$ and $\theta$ representing the density fluctuations
in the spin-up state.
Similar to Refs.~[\onlinecite{adilet,parola,affleck}], we present the effective ``quantum
impurity'' Hamiltonian in the form
\begin{eqnarray}
H=H_0&+&\int\! dx\, d^\dagger\, [\omega_m(q,n)-v_m(q)(-i\partial_x )]\, d\nonumber\\
&+&\int\! dx\, [V_\phi(\partial_x\phi)
        +V_\theta(\partial_x\theta)]\, d^\dagger d\, ,
\label{qimp}
\end{eqnarray}
where $v_m(q)=\partial_q\omega_m(q,n)$. The first of the two added
parts to $H_0$ establishes $\omega_m(q,n)$ as the lower edge for the
spectral function~\cite{footnote}
\begin{equation*}
A_m(q,\omega)=\sum_\nu\delta(\omega-E_\nu(q))|\langle\nu,q|s_-(q)|0\rangle|^2\,,
\end{equation*}
where $|\nu,q\rangle$ is a many-body eigenstate of the system with
momentum $q$ and energy $E_\nu(q)$.  The gradient expansion used in
that part is sufficient for finding the behavior
of $A_m(q,\omega)$ in the vicinity of the edge.
The last term in the r.h.s. of Eq.~(\ref{qimp}) describes interactions
of the quantum impurity with the density waves. The strengths of such
interactions, $V_\phi$ and $V_\theta$, can be expressed through
independently measurable characteristics of the system.

To determine $V_\theta$, we note that the corresponding term in the
Hamiltonian Eq.~(\ref{qimp}) is nothing but a modification of the
energy-momentum relation, $\omega_m^u(q,n)=\omega_m(q)+V_\theta
m_\uparrow u$, for the $d$-quasiparticle (a mobile ``impurity'') in
the presence of finite velocity $u=\partial_x\theta/m_\uparrow$ of the
fluid formed by the spin-up particles.  That allows~\cite{baym} one to
use the Galilean invariance to find $V_\theta$. Indeed, in the
presence of a flow of the liquid with velocity $u$, the magnon energy
remains unchanged in the co-moving frame. In the laboratory frame, the
magnon momentum is $q+m_\downarrow u$, while its energy is
$\omega_m^u(q+m_\downarrow u,n)=\omega_m(q,n)+qu+m_\downarrow u^2/2$.
Changing the momentum in the last formula, $q+m_\downarrow u\to q$,
and comparing the two above expressions for $\omega_m^u(q,n)$, we
find in the limit of small $u$:
\begin{equation}
V_\theta(q)
=\frac{q}{m_\uparrow}-\frac{m_\downarrow}{m_\uparrow}\,\partial_q\omega_m(q,n).
\label{baym}
\end{equation}
Hereinafter we assume the mass of bosons is  spin-independent,
$m_\uparrow=m_\downarrow=m$, thus reducing Eq.~(\ref{baym}) to
\begin{equation}
V_\theta(q)=\frac{q}{m}-v_m(q).
\label{vtheta}
\end{equation}
At $|q|\ll \pi n$, the magnon spectrum is quadratic in $q$, 
i.e. $\omega_(q)\approx \omega_m(0)+q^2/(2m^*)$ with
some effective mass $m^*$; in this limit, $V_\theta=[(m^*-m)/m]v_m$.

To determine $V_\phi$, we may consider the effect of a long wavelength
density variation, $\delta\rho=(1/\pi)\nabla\phi$ on the
energy of the system. According to Eq.~(\ref{qimp}), it adds a term
$V_\phi\pi\delta\rho$ to the energy density. This variation should be
equal to the corresponding value,
$(\partial\omega_m/\partial n)+(\partial\mu/\partial n)$ defined
phenomenologically (here $\mu$ is the chemical potential of the
spin-up bosons). Expressing $\partial\mu/\partial n$ in terms of
$K$, one finds~\cite{adilet2}
\begin{equation}
V_\phi(q)=\frac{1}{\pi}\frac{\partial\omega_m(q,n)}{\partial n}+\frac{v}{K}.
\label{vphi}
\end{equation}
As follows from the time-reversal symmetry of the problem, $V_\theta(q)$ and
$V_\phi(q)$ are correspondingly odd and even functions of the momentum $q$.

It is instructive to consider separately the $q=0$ limit.  An
introduction of a static impurity (spin-down boson with $q=0$) creates
displacement $-(\pi KV_\phi/v)\sign (y-x)$ in field $\phi$, while
operator $\exp[i\theta (x)]$ in the definition of the spin density
Eq.~(\ref{sminus}) creates a shift $\pi\sign (y-x)$. The sum of the
two is related to the dilatation $\delta_l$ caused by the spin flip,
$-\pi KV_\phi/v+\pi=\pi\delta_l$, yielding $V_\phi=(v/K)(1-\delta_l)$.
Comparing that with Eq.~(\ref{vphi}) at $q=0$, we find
\[
\delta_l=-\frac{K}{\pi v}\frac{\partial\omega_m(0,n)}{\partial n}\,.
\]
Derived from considering a single spin flip, this relation under some
assumptions can also be derived from the thermodynamics of a system at
a constant pressure. One must assume that magnons do not form bound
states, so $\omega_m(0,n)=\partial{\cal E}/{\partial S}$ with ${\cal
  E}$ being the ground-state energy and $S$ the total spin. In the
presence of SU(2) invariance $\partial{\cal E}/{\partial S}=0$, thus
flipping a spin with $q=0$ does not cause dilatation, $\delta_l=0$.

Once the interaction constants $V_\theta$ and $V_\phi$ are
established, we may proceed in full analogy with
Ref.~[\onlinecite{adilet2}]: (i) re-scale variables,
$\phi=\tilde{\phi}\sqrt{K}$, $\theta=\tilde{\theta}/\sqrt{K}$ thus
giving Eq.~(\ref{liquid}) the appearance of a Hamiltonian of free
``particles''; (ii) eliminate the linear in $\phi$, $\theta$ part of
the full Hamiltonian, Eqs.~(\ref{liquid}) and (\ref{qimp}), by a
unitary transformation with proper values of $\delta_+$, $\delta_-$,
\begin{equation}
U^\dagger=e^{-i\int dx
\left\{\frac{\delta_+(q)}{2\pi}[\tilde{\theta}(x)-\tilde{\phi}(x)]-
\frac{\delta_-(x)}{2\pi}[\tilde{\theta}(x)+\tilde{\phi}(x)]\right\}
d(x)d^\dagger(x)}\, ;
\label{U}
\end{equation}
(iii) express the exponents of
the sought correlation functions in terms of the phase shifts
$\delta_+$ and $\delta_-$ which ``quantum impurity''
($d^\dagger d=1$) causes for co-moving ($+$) and counter-propagating
($-$) ``particles''.
The values of the phase shifts can be written in
a relatively compact form as:
\begin{eqnarray}
\frac{\delta_\pm(q)}{\pi}
&=&\frac{1}{v_m(q)\mp v}\left(\sqrt{K}\, V_\phi \pm \frac{1}{\sqrt{K}}\, V_\theta \right)  .
\label{deltas}
\end{eqnarray}

We are now in the position to evaluate the correlation functions of
interest. Using Eq.~(\ref{sminus}) we may represent $A_m(x,t)$ as
\begin{eqnarray}
&&A_m(q,\omega)\label{Asxt}\\
&&=\Im \int \!dxdt\, e^{iqx-i\omega t}
\langle 0| d(0,0) e^{-i\theta(0,0)}e^{i\theta(x,t)}
    d^\dagger(x,t)|0\rangle.
\nonumber
\end{eqnarray}
In the case of spinor Bose liquid one may envision tunneling of a
spin-up or a spin-down particle into otherwise fully spin-up polarized
system. The spectral function for a spin-up particle is identical to
the one evaluated for the liquid of spinless
bosons~\cite{khodas,adilet}; the corresponding tunneling theshold
spectrum is $\omega=vq$. For a spin-down particle, the tunneling
threshold is determined by the magnon spectrum, and the tunneling
probability is proportional to the spectral function of the $d$-quasiparticle,
\begin{equation}
A_d(q,\omega)=\Im \int \!dxdt\, e^{iqx-i\omega t}
\, \langle 0|d(x,t) d^\dagger(0,0)|0\rangle\,.
\label{Adxt}
\end{equation}
Functions $A_m(q,\omega)$ and $A_d(q,\omega)$
exhibit power-law behavior above the threshold,
\begin{equation*}
    A_{m,d}(q,\omega)\propto \Theta\big[\omega-\omega_m(q)\big]
    \big[\omega-\omega_m(q)\big]^{\mu_{m,d}(q)}\,.
\end{equation*}
Using the transformation Eq.~(\ref{U}) and the standard methods of
bosonisation, one finds for $\mu_d(q)$ in the region $|q|\leq\pi n$:
\begin{equation}
\mu_d(q)=-1+\left(\frac{\delta_+(q)}{2\pi}\right)^2
+\left(\frac{\delta_-(q)}{2\pi}\right)^2,
\label{mud}
\end{equation}
like in Ref.~[\onlinecite{adilet2}]. Similarly, the exponent
$\mu_m(q)$ at $|q|\leq\pi n$ reads as
\begin{equation}
\mu_m(q)=-1+\left({1\over 2\sqrt{K}}+ \frac{\delta_+(q)}{2\pi}\right)^2
+\left({1\over 2\sqrt{K}} -\frac{\delta_-(q)}{2\pi}\right)^2.
\label{mus}
\end{equation}
Equations (\ref{mud}) and (\ref{mus})
together with Eq.~(\ref{deltas}) relate the exponents of the
correltaion functions to the properties of the magnon branch of
excitation spectrum  $\omega_m(q,n)$ for the principal interval of momenta
$|q|\leq \pi n$.

Due to the peculiarity of 1D systems, the lowest-energy excitations
corresponding to a single flipped spin or to an added spin-down
particle at a given momentum $q$, are periodic functions of the
momentum, $\omega_m(q,n)=\omega(q-2\pi nl, n)$ for any integer $l$. To
extend the above results for $\mu_{m,d}(q)$ beyond the principal
interval of momenta, we notice that introducing a $d$-quasiparticle
with the lowest energy amounts to a momentum boost $2\pi n l$ of the
spin-up liquid and exciting a magnon with the residual momentum
belonging to the principal interval $-\pi n<q-2\pi nl\leq \pi n$. The
boost accompanying the creation of a $d$-quasiparticle corresponds to
the modified definition of the spin-down particle creation operator:
\begin{equation}
d^\dagger(x) \to e^{-2il\phi(x)}d^\dagger(x)\,;\quad \quad
d(x) \to d(x) e^{2il\phi(x)}\,.
\label{psi}
\end{equation}
Performing these replacements in Eq.~(\ref{Adxt}) and repeating the steps which have lead to
Eq.~(\ref{mud}), we find now:
\begin{equation}
\mu_d(q)=-1+\left(\frac{\delta_+(q^*)}{2\pi}-l\sqrt{K}\right)^2
+\left(\frac{\delta_-(q^*)}{2\pi}-l\sqrt{K}\right)^2,
\label{mudl}
\end{equation}
where $q^* = q-2\pi nl$ with the integer $l$ chosen to have
$|q^*|\leq\pi n$. A similar procedure for $A_m(q,\omega)$ yields
\begin{eqnarray}
\mu_m(q)=-1&+&\left({1\over 2\sqrt{K}}+ \frac{\delta_+(q^*)}{2\pi}-l\sqrt{K}\right)^2
\nonumber \\
&+&\left({1\over 2\sqrt{K}} -\frac{\delta_-(q^*)}{2\pi}+l\sqrt{K} \right)^2.
\label{musl}
\end{eqnarray}
Equations (\ref{mudl}), (\ref{musl}) along with Eq.~(\ref{deltas})
provide the values of the edge exponents for an arbitrary momentum $q$.

Transferring momentum to the liquid as a whole allows tunneling of a
particle at low energy, $\omega\to\omega_m(q^*,n)$, even at high
momentum $|q|>\pi n$. The price for that, however, is a reduced
tunneling probability reflected by the presence of integer $l$ in
Eq.~(\ref{mudl}): while the spectrum $\omega_m(q)$ is periodic, the
exponent $\mu_d(q)$ is increasing with moving from one period to
another, with larger $|l|$. The suppressed tunneling probability is a
manifestation of the orthogonality catastrophe. Similarly the exponent
$\mu_m$, describing the probability of the spin-flip photon absorption
near the edge, is increased due to the orthogonality catastrophe.
Indeed, for $|l|>0$ the final state includes the spin-flipped particle
along with the moving fluid with the momentum $2\pi nl$, which has the
progressively smaller overlap with the initial state of the fluid at rest.

The periodic dispersion relation reaches its maxima at $q=\pi
n(2l-1)$.  Depending on the microscopic interaction strength between
the bosons, the magnon velocity $v_m=\partial_q\omega_m(q,n)$ may have
jumps at $q=\pi n (2l-1)$, or be a continuous function. In the latter
case, obviously, $v_m(\pi n (2l-1))=0$. The transition between the two
types of behavior upon the increase of the interaction strength is
equivalent to the ``quantum phase transition'' in the Kondo problem
controlled by tuning the exchange constant through
zero~\cite{lamacraft}. The $v_m(\pi n (2l-1))=0$ regime corresponds to
the strong-coupling side of the transition. The developed above
Luttinger liquid representation is applicable on either side of the
transition, similar to the scatterling phase description of the
low-energy physics of the Kondo problem.  The region of applicability
in $\omega-\omega_m(q,n)$, of course, gets narrow close to the
transition point, as the corresponding Kondo energy scale becomes
small. In the strong-coupling regime, $v_m$ and
$\partial_n\omega_m(q,n)$ have no discontinuities at $q=\pi n(2l-1)$.
It is not clear {\it apriori} that the same is true for the exponents
$\mu_{m,d}(q)$: after all, the definition of the response functions
involves different operators, see Eq.~(\ref{psi}), at subsequent
intervals of momenta. It is quite striking to see directly from
Eqs.~(\ref{mudl}), (\ref{musl}) and (\ref{deltas}) the continuity of
$\mu_d(q)$ and $\mu_m(q)$.  Indeed, substitution of $v_m(q^*=\pm\pi
n)=0$, the use of relations $K=\pi n/(mv)$ and
$\partial_n\omega_m(q^*,n)|_{q^*=\pi
  n}=\partial_n\omega_m(q^*,n)|_{q^*=-\pi n}$ in Eq.~(\ref{deltas})
yields:
\begin{equation}
\frac{\delta_\pm(\pi n)}{2\pi} = \frac{\delta_\pm(-\pi n)}{2\pi}
-\sqrt{K}\,.
\label{plus-minus-pin}
\end{equation}
With the help of Eqs.~(\ref{mudl}) and (\ref{musl}) this immediately
implies that for $q\to \pi n (2l-1)\pm 0$ both
$\mu_d(q)$ and $\mu_m(q)$ are continuous functions {\em and} their first
derivatives over $q$ are continuous as well.  At the ``weak-coupling''
  side of the transition, the exponents, together with $v_m(q)$, are
  discontinuous at $q=\pi n (2l-1)$.

Around the local minima, $q= 2\pi nl$, i.e. at $q^*= 0$, one finds
\begin{equation}\label{mud1}
  \mu_d(2\pi n l)=-1+\frac{(1-\delta_l)^2}{2K} + 2l^2 K\,
\end{equation}
and
\begin{equation}\label{mus1}
    \mu_m(2\pi nl)=-1 +\frac{\delta_l^2}{2K}+2l^2K\,.
\end{equation}
The fact that $\mu_{d}(q=0)\neq -1$ is due to the orthogonality
catastrophe\cite{anderson}: the tunnelled spin-down boson shakes up
the liquid of spin-up particles. The same mechanism have
led\cite{adilet} to non-trivial tunneling exponents
$\overline{\mu_\pm}$ for spinless bosons in the
Lieb-Liniger\cite{lieb} model, once the exponents are evaluated beyond
the Luttinger liquid approximation. The direct comparison of
Eq.~(\ref{mud1}) with the corresponding result, Eq.~(22) of
Ref.~[\onlinecite{adilet}] is possible only at $K\to 1$, when the
shake-up becomes independent of the impurity velocity relative to $\pm
v$; the two equations agree with each other, yielding
$\mu_d=\overline{\mu_\pm}=1/2$. The similar physics is at work for
$A_m(q,\omega)$ correlation function away from the $SU(2)$ symmetric
point, i.e. when $\delta_l\neq 0$. The amplitude of a spin-flip
process for a finite dilatation $\delta_l$ is suppressed by the
orthogonality, resulting in $\mu_m(0)\neq -1$.

In the $SU(2)$ invariant case, where $\partial\omega_m(0,n)/\partial
n\propto \delta_l=0$, one may also deduce universal results for the
momentum dependence of $\mu_{m,d}(q)$ in the region of small momenta
$|q|\ll m^*v$,
\begin{equation}\label{mud2}
  \mu_d(q)=-1+\frac{1}{2K}
+ \frac{K q^2}{2(\pi n)^2}\left( 1 +{m\over m^*}\,(2+2\sigma) \right)\, ,
\end{equation}
where $\sigma= -(n/2m^*)\partial  m^*/\partial n$. Similarly,
\begin{equation}\label{mus2}
    \mu_m(q)=-1
+ \frac{K q^2}{2(\pi n)^2}\left[ 1 + \left(\frac{q}{m^*v}\right)^2
\left(3+4\sigma +\sigma^2 \right)\right]\, .
\end{equation}
The $q^2$ term of the $\mu_m(q)$ exponent depends only on the
parameters of the Luttinger liquid and for the strong coupling limit
was derived in Refs.~\onlinecite{zvon,akirako}.  It is interesting
that the $q^4$ term here may be expressed through the effective mass
of the magnon mode $m^*$. The latter was evaluated in various limits
for the integrable contact-interaction model~\cite{fuchs}, which
allows us finding $\sigma$ in Eqs.~(\ref{mud2}) and (\ref{mus2}). In
the strong coupling limit $m^*=3\gamma m/(2\pi ^2)$, where $\gamma =
mg/n\gg 1$ and thus $\sigma=1/2$; here $g$ is the interaction
strength.  In the weak coupling limit $m^*=m(1+2\sqrt{\gamma}/3\pi)$
and thus $\sigma =\sqrt{\gamma}/6\pi\ll 1$. These considerations are
also applicable for the $q^2$ term in $\mu_d(q)$.

In conclusion, the dynamic response functions Eqs.~(\ref{Asxt}),
(\ref{Adxt}) of a homogeneous ferromagnetic one-dimensional Bose
liquid exhibit power-law asymptotes at the threshold defined by the
spectrum of the magnon $\omega_m(q)$. Independent of any model
assumptions, the corresponding exponents $\mu_m(q)$ and $\mu_d(q)$ at
any wave vector can be expressed in terms of a few independently
measurable parameters: the sound velocity $v$, the corresponding
Luttinger liquid parameter $K$, and the derivatives of $\omega_m$ with
respect to $q$ and liquid density $n$. Further simplification of
$\mu_m(q)$ and $\mu_d(q)$ is possible in the vicinities $q=2\pi n l$
of the minima of magnon spectrum.

We thank D. Gangardt, K.A. Matveev and A. Lamacraft for useful  discussions.  This research is
supported by DOE Grant No. DE-FG02-08ER46482.


\begin{thebibliography}{99}
\bibitem{footnote} The correlation functions of various spin
  components form a tensor. For the ferromagnetic ground state, $\langle
  s_+(x,t)s_-(0,0)\rangle$ is its only non-trivial component.
\end{thebibliography}
\end{document}